\documentstyle[aps]
              {revtex}
\begin{document}
\draft
\title{
Internal thermal noise in the LIGO test masses :
a direct approach.
}
\author{  Yu. Levin
}
\address{Theoretical Astrophysics, California Institute of
Technology, Pasadena, California 91125}
\date{\today}
\maketitle
\begin{abstract}
The internal thermal noise in LIGO's test masses is analyzed by a new technique,
a direct application of the Fluctuation-Dissipation Theorem to LIGO's readout observable,
$x(t)=$(longitudinal position of test-mass face, weighted by laser beam's Gaussian profile).
Previous analyses, which relied on a normal-mode decomposition of the test-mass motion,
were valid only if the dissipation is uniformally distributed over the test-mass interior, and
they converged reliably to a final answer only when the beam size was a non-negligible fraction
of the test-mass cross section. This paper's direct analysis, by contrast, can handle inhomogeneous 
dissipation and arbitrary beam sizes. In the domain of validity of the previous analysis, the two methods
give the same answer for $S_x(f)$, the spectral density of thermal 
noise, to within expected accuracy. The new analysis predicts that thermal noise
due to dissipation concentrated in the test mass's front face (e.g. due to mirror coating)
scales as $1/r_0^2$, by contrast with homogeneous dissipation, which scales as $1/r_0$ ($r_0$ is the beam radius);
so surface dissipation could become significant for small beam sizes. 
\end{abstract}

\narrowtext
\section{introduction}

Random thermal fluctuations are expected to be the dominant noise source
for the first interferometers in the Laser Interferometer Gravitational Wave Observatory
(LIGO) at  frequencies between $35$ and $100$ Hz. This thermal noise is generally decomposed 
into a suspension thermal noise and an internal thermal noise for the test masses. The former can be traced
back to the friction in  the test masses' pendular  suspension system; the latter is due to  internal damping
inside the test masses themselves.  Traditionally, themal noise calculations have been based on a 
  normal-mode expansion  \cite{saulson},\cite{raab} 
. However, Gonzalez and Saulson have also
performed an exact calculation of the suspension thermal noise by applying directly
the Fluctuation-Dissipation (FD) theorem \cite{gonzales} in it's most general form,
due to H.\ B.\ Callan and T.\ A.\ Welton \cite{callen}. The purpose of this paper
is to use the general method of Gonzales and Saulson 
to calculate the internal thermal noise.

In Section II we analyze a general situation when a measuring device (e.g. a laser interferometer)
monitors the displacement of the surface of a test mass whose internal degrees of freedom
are in themal equilibrium with each other. We develop a general 
formalism for using the FD theorem to calculate the thermal noise in the most general surface 
readout quantity.
In brief our method is as follows:

To work out the thermal noise at a particular frequency $f$, one should mentally apply  pressure 
oscillating at this frequency 
to the observed surface of the test mass. 
  The spatial variation of this pressure should mimic that of the light beam
intensity (for example, in the case of a gaussian beam this oscillating pressure has a
gaussian profile of the same widthi as the beam). The thermal noise is then given by
\begin{equation}
S_x(f)={2 k_{B} T\over \pi^2 f^2} {W_{diss}\over F_0^2},
\label{eq:tn1}
\end{equation}
where $k_{B}$ and $T$ are the Boltzmann's constant and the temperature of
the mirror respectively, 
$F_0$ is the amplitude of the oscillating force applied to the surface (i.e.  
 the pressure integrated over the surface), and  $W_{diss}$ is the 
time-averaged power dissipated in the test mass when this
oscillating pressure is applied.     
 
To demonstrate the computational power of this general approach, in Section III we consider
the case of a cylindrical fused silica test mass monitored by a circular gaussian
laser beam. For the case when the radius  of the beam is much less then the size of the test mass
and the dissipation is uniformly distributed throughout test mass volume,
we derive an analytical expression for the thermal noise [cf. Eq. (\ref{eq:tn3}) of Section III
]:
\begin{equation}
S_x(f)={4k_{B} T\over f}{1-\sigma^2\over \pi^3 E_0 r_0} I \phi\left[1+O\left({r_0\over R}\right)\right].
\label{eq:tn2}
\end{equation}
Here $\sigma$, $E_0$, and $\phi$ are the Poisson ratio, Young's modulus, and dissipational loss
angle [Eq. (\ref{eq:ym})] of the test-mass material, $r_0$ is the radius
of the laser beam (which is defined here as a radius at which the intensity of light is $1/e$ of the maximum
intensity)
, $R$ is a characteristic size of the test
mass, and $I=1.87322...$ in the case of  a gaussian beam. Putting numbers in Eqs. (\ref{eq:tn1}) and (\ref{eq:tn2}),
we find that our results are in agreement with those of Raab and Gillespie \cite{raab}, who used the more
complicated and computationally involved method of normal-mode decomposition. It is interesting to note
that as $r_0/R$ tends to zero, our simple analytical formula becomes more precise, whereas the 
more complicated and computationally involved method of normal-mode decomposition requires summing 
over a larger number of modes and thus becomes computationally more expensive.

Not only can the normal-mode decomposition  be computationally expensive, it can also be misleading.
We demonstrate this point in Section IV by considering a test mass which has a lossy surface, e.g. 
due to  a lossy mirror coating.
We estimate the contribution of the surface to the thermal noise using the general
method of Section II, and show that it differs from
the estimate obtained by the method of normal modes (which gives a result too small by a factor
of at least $\sim r_0/R$).
 This breakdown of the normal-mode analysis 
will in general happen   when the sources of friction are not distributed homogeneously
over the test mass. The fundamental reason  is that in this case different normal modes
can have a common Langevin driving force  (which is not so if the defects are
distributed homogeneously).

Our analysis shows that thermal noise due to surface losses near the laser beam spot 
scales as $S_x(f)\propto 1/r_0^2$, whereas thermal noise due to volume losses scales as $1/r_0$.
Correspondingly, for small beam spots the surface losses could become significant. To protect against
this, it is important to keep the surface near the laser beam spot as free of potential sources of friction as 
possible.

\section{General method}
For concreteness, consider a situation where LIGO's laser beam is shining
on the  circular surface  of one of LIGO's cylindrical test masses.
The phase shift of the reflected light contains information about the motion of
the test mass's surface. The  variable read out by this procedure can be written
as 
\begin{equation}
x(t)=\int f(\vec{r}) y(\vec{r},t) d^2 r.
\label{eq:readout}
\end{equation}
Here $\vec{r}$ is the transverse location of a point on the 
test-mass surface, and $y(\vec{r},t)$ is the 
displacement of the boundary along the direction of the laser beam at point $\vec{r}$
and time $t$. The form factor $f(\vec{r})$ depends on the laser beam profile and is proportional
to the laser light intensity at the point $\vec{r}$ \cite{raab}; it is normalized by 
$\int f(\vec{r}) d^2r =1$.

The internal thermal noise of the test mass is defined as the fluctuations
in $x(t)$, and our objective is to find the spectral density $S_x(f)$ of these 
fluctuations. We assume that the test mass is in thermal equilibrium at  temperature
$T$.

Callen and Welton's generalized Fluctuation-Dissipation Theorem 
\cite{callen} says that the spectral density
of the fluctuations of LIGO's readout variable $x(t)$ is given by the formula
\begin{equation}
S_x(f)={k_{B} T\over \pi^2 f^2}\left|Re\left[Z\left(f\right)\right]\right|,
\label{eq:FD1}
\end{equation}
where $k_B$ is Boltzman's constant and $Z(f)$ is a complex impedance
associated with $x(t)$. This complex impedance can be understood and 
computed as follows. Introduce a special set of generalized coordinates for the test mass's
degrees of freedom--a set for which $x$ is one of the coordinates. (Since $x$ is not the coordinate 
of a normal mode of the test mass, these generalized coordinates will not be the usual ones 
associated with normal modes.) Apply to the test mass a generalized force $F(t)$ that drives the generalized
 momentum conjugate to $x$ but does not drive any of the other generalized momenta. This generalized force 
will show up as the following interaction term in the test mass's Hamiltonian:
 \begin{equation}
H_{\rm int}=-F(t) {x}.
\label{eq:hamiltonian}
\end{equation} 
This driving force, together with the test mass's internal elastic forces and internal dissipation,
will generate a time evolution $x(t)$ of the observable $x$. Denote by $F(f)$ and $x(f)$ the Fourier transforms
of the (arbitrary) driving force $F(t)$ and the observable's response $x(t)$. Then the impedance that appears
in the thermal noise formula Eq. (\ref{eq:FD1}) is
\begin{equation}
Z(f)=2\pi\imath f x(f)/F(f).
\label{eq:impedance}
\end{equation}

The physical nature of the driving force $F(t)$ can be deduced by inserting the 
definition (\ref{eq:readout}) of the observable $x$ into the interaction Hamiltonian
(\ref{eq:hamiltonian}):
\begin{equation}
H_{\rm int}=-\int P(\vec{r})y(\vec{r},t)d^2 r,
\label{eq:hamiltonian1}
\end{equation}
where
\begin{equation}
P(\vec{r},t)=F(t) f(\vec{r}).
\label{eq:pressure}
\end{equation}
From Eq. (\ref{eq:hamiltonian1}) we see that the generalized force $F(t)$ consists of
a pressure $P(\vec{r},t)$ [Eq. (\ref{eq:pressure})] applied to the test mass's surface.
Note that the spatial distribution of this pressure is the same as LIGO's laser beam intensity
profile.

The real part of the impedance, $Re[Z(f)]$, describes the coupling of the test mass's dissipation
to the observable $x$. We can see this most clearly by applying an oscillatory
pressure $P(\vec{r},t)=F_0\cos(2\pi ft)f(\vec{r})$ to the test mass's face. From the
response formula (\ref{eq:impedance}) we infer that the power $W_{\rm diss}$ that this oscillatory pressure feeds
into the test mass, and that the test mass then dissipates, is related to $|Re[Z(f)]|$ by
\begin{equation}
\left|Re\left[Z\left(f\right)\right]\right|={2W_{\rm diss}\over F_0^2}.
\label{eq:impedance1}
\end{equation}
Substituting Eq. (\ref{eq:impedance1}) into Eq. (\ref{eq:FD1}), we get
\begin{equation}
S_x(f)={2k_B T\over \pi^2 f^2} {W_{\rm diss}\over F_0^2}.
\label{eq:FD2}
\end{equation}

Equation (\ref{eq:FD2}) is the most important equation of this paper.
Let us reephasize it's physical content:\newline
\begin{enumerate}
\item  Apply an oscillatory pressure \newline
$P(\vec{r},t)=F_0 \cos (2\pi ft) f(\vec{r})$
to the face of the test mass.\newline
 \item Work out the average power $W_{\rm diss}$ dissipated in the test mass
under the action of this oscillatory pressure.\newline
 \item Use $F_0$ and $W_{\rm diss}$ in Eq. (\ref{eq:FD2}) to calculate
$S_x(f)$.\newline
\end{enumerate}

This procedure is different from the one employed in  previous calculations of 
 internal thermal noise for the LIGO and VIRGO test masses 
\cite{saulson}, \cite{raab}, \cite{vinet}. The previous authors
decomposed a  test mass's motion into normal elastic modes; then they calculated 
the contribution of each mode to  $S_x$ independently and added
 up these contributions.
This method of ``normal-mode decomposition'' works fine 
in many cases, but it has two drawbacks:\newline
\begin{enumerate}
\item  The fundamental assumption in this method is that different normal modes have 
independent   Langevin forces. This assumption is correct only if the 
sources of friction are homogeneously distributed over the  test-mass volume.
It breaks down if the defects are more concentrated in one place than in others---
for example, when there is  significant damping concentrated in the
test-mass surface. We will return to this in  
Section IV.\newline
\item For a small laser beam diameter the sum over normal modes converges very
slowly, so
one has to sum over many modes, which may be computationally expensive.
By contrast, using the new method described in this paper, one can write down a
simple analytic expression for the low-frequency noise in the case of a narrow laser beam.
In the next section we derive this expression and make comparison with the normal-mode decomposition
results derived in  \cite{raab}.
\end{enumerate}

\section{thermal noise due to homogeneously distributed damping}
Consider the case where all the friction in the test mass comes from  homogeneously
distributed damping.
It is conventional to characterize such friction by an imaginary part of 
the material's Young's modulus:
\begin{equation}
E=E_0\left[1+\imath\phi (f)\right];
\label{eq:ym}
\end{equation}
$\phi(f)$ is called the material's ``loss angle''.
It is suspected \cite{raab1}, \cite{saulson} that for fused silica, which will be used in
LIGO's test masses, $\phi$ might be independent of frequency within LIGO's detection band
(but there is no evidence for such behavior of $\phi$ for figh-quality 
resonators---see \cite{braginsky} for some healthy scepticism).
In this $f$-independent case the damping is called ``structural''.

To calculate the thermal noise for homogeneous dissipation, 
we express $W_{\rm diss}$ in Eq. (\ref{eq:FD2}) as
\begin{equation}
W_{\rm diss}=2\pi f U_{\rm max}\phi(f),
\label{eq:wdiss1}
\end{equation} 
where $U_{\rm max}$ is the energy of elastic deformation at a moment when the test mass is 
maximally contracted or extended
 under the action of the oscillatory pressure 
of Eq. (\ref{eq:pressure}).

LIGO's detection frequencies  ($10-300$Hz) are much lower than the eigenfrequencies
of the test mass's normal modes (the lowest of which is $\sim 6$kHz); so we can assume
constant, non-oscillating pressure $P(\vec{r})=F_0 f(\vec{r})$ when evaluating $U_{\rm max}$.

In the case when the beam profile is gaussian and the centre of the light spot coincides with the 
centre of the transverse coordinates, we have
\begin{equation}
f(\vec{r})={1\over \pi r_0^2}e^{-r^2/r_0^2},
\label{eq:formfactor}
\end{equation}
where $r_0$ is the radius of the laser beam. When the characteristic size of the test mass
$R$ is much greater than $r_0$, we can approximate the test mass as 
an infinite half-space in order to find $U_{\rm max}$. Appendix A uses elasticity theory to derive 
$U_{\rm max}$ in this  case [cf. Eq. (\ref{eq:umax2})]:
\begin{equation}
U_{\rm max}= {F_0^2\over\pi^2 E_0 r_0}(1-\sigma^2)I\left[1+O\left({r_0\over R}\right)\right],
\label{eq:umax1}
\end{equation}
where $E_0$ and $\sigma$ are the Young's modulus and  Poisson ratio of the material respectively,
and $I\simeq 1.87322$. Here $O(r_0/R)$ is a correction due to the finite size
of the cylinder. 
Putting Eqs. (\ref{eq:umax1}) and (\ref{eq:wdiss1}) into Eq. (\ref{eq:FD2}),
one gets
\begin{equation}
S_x(f)={4k_{B} T\over f}{1-\sigma^2\over \pi^3 E_0 r_0} I \phi\left[1+O\left({r_0\over R}\right)\right].
\label{eq:tn3}
\end{equation}

 Below
we take the numerical values \footnote{Note that our definition
of the beam radius (location where intensity has fallen to $1/e$ of its central
value) differs by $\sqrt{2}$ from the beam radius of Ref. \cite{raab} (location
of $1/e$ amplitude falloff).}  used by
Gillespie and Raab \cite{raab}:
  $r_0=1.56\hbox{cm}$,  $E_0=7.18\times 10^{10}\hbox{Pa}$, $\sigma=0.16$, $\phi=10^{-7}$,
the mirror diameter of $25\hbox{cm}$ and the mirror length of $10\hbox{cm}$. Gillespie and
Raab, after summing over the relevant $\sim 30$ modes, get
 \begin{equation}
S_x^{\rm GR}(100\hbox{Hz})\simeq 8.0\times 10^{-40}\hbox{m$^2$/Hz}.
\label{eq:modes}
\end{equation}
Our analytical approximation (\ref{eq:tn3}) (which should be  valid to within $\sim 10$ percent in this case)
gives 
\begin{equation}
S_x(100\hbox{Hz})\simeq 8.7\times 10^{-40}\hbox{m$^2$/Hz}.
\label{eq:numerics}
\end{equation}
Notice that our analytic expression in Eq. (\ref{eq:tn3}) gets more exact when $r_0/R\rightarrow 0$,
whereas, by contrast, the sum over modes converges more slowly and gets more complicated.

The ratio $r_0/R$ may turn out to be of order unity in real experiments. In this case,
Eq. (\ref{eq:tn3}) can only be used for order-of-magnitude estimates. To work out the exact value
of the internal thermal noise, one would need to calculate $U_{\rm max}$ numerically.
We have done such a numerical computetion using finite-element techniques. More 
specifically, we have used finite-element software called PDEase2D [Version 3.0], which runs as part of
Mascyma [Version 2.1], 
 to  solve the elasticity equations for
the loaded mirror and to compute $U_{\rm max}$ and, by virtue of Eqs. (\ref{eq:wdiss1}) and
(\ref{eq:FD2}), $S_x$.
The exact answer for the mirror and light spot parameters given above is
\begin{equation}
S_x(100\hbox{Hz})=8.76\times 10^{-40}\hbox{m$^2$/Hz},
\label{eq:exact}
\end{equation}
which is consistent (better than expected) with our analytical approximation.

The purpose of the present section is  to convince the reader that the method presented
in this paper is correct and could be computationally cheaper than the normal-mode expansion.
The next section concentrates on the cases where a direct application of the FD theorem 
can be crucial for getting the right results, and the method of normal-mode decomposition
fails.

\section{ the case of surface damping}
In this section we study thermal noise due to surface losses--caused, e.g., by
inadequate polishing or by a lossy mirror coating.

From Eq.\ (\ref{eq:FD2}) we see that the key quantity in the thermal noise calculation
is the power dissipated in the test mass when an oscillating pressure is applied to the laser
beam spot on the test-mass surface. The power dissipated at each  point of the material
is proportional to the square of the stress at this point. Most of the surface stress is in or
near the spot to which the pressure is applied, so 
\begin{equation}
W_{\rm diss}^{\rm coating}\propto \left({F_0\over r_0^2}\right)^2 r_0^2={F_0^2\over
r_0^2}.
\label{eq:wdiss2}
\end{equation} 
Thus the thermal noise due to the surface damping scales like
\begin{equation}
S_x(\hbox{boundary})\propto 1/r_0^2.
\label{eq:boundary}
\end{equation}
For comparison, the thermal noise due to bulk damping
[Eq. (\ref{eq:tn3})] scales as
\begin{equation}
S_x(\hbox{bulk})\propto 1/r_0.
\label{eq:bulk}
\end{equation}
Thus as the spot size decreases, the thermal noise due to surface damping
grows faster than that due to  bulk damping.

Contrast this conclusion with the intuition one gets from normal-mode decomposition.
There one is concerned with how much  the surface contributes to 
the quality factors ($Q$'s) of the 
normal modes. For a typical mode the strain at the surface is at most
of the same order as the characteristic strain inside the test mass
(likely, much less for first few  modes---because of the free boundary condition).
Therefore, one would presume that the surface contributes no more
than some mode-independent fraction   of the test mass's $Q$'s.
In order of magnitude this fraction should be the ratio of the power
dissipated in the surface to that in the bulk 
if one applies an oscillating
pressure uniformally to the whole surface, which in the context of our
method corresponds to a beam radius of $R$. Therefore the normal-mode
 estimate of the surface thermal noise
is at least $r_0/R$ less than the correct value. 

Current experiments show that the mirror
coating does not contribute significantly to the $Q's$ of the test-mass normal modes.
 The conclusion commonly made is  that coating is also not likely to contribute significantly
to the internal thermal noise. The above analysis shows that
this conclusion is not justified and that there might be a significant 
contribution of the coating to the internal thermal noise, despite the fact that $Q$'s are not
significantly changed.

\section{ discussion and conclusion}

The normal-mode decomposition of the thermal noise is exact
when the defects are distributed homogeneously through 
the volume of the test mass. However, as was shown explicitly 
in Section IV for 
the case of surface losses, when the defect
distribution is not homogeneous, the normal-mode decomposition 
may be misleading, and a direct application  of the
Fluctuation-Dissipation theorem is required. 

Thermal noise is ultimately linked to  friction
in the test mass; this friction is caused by various
(structural and otherwise) defects. Those defects 
which are closer to the beam spot will contribute
more to the thermal noise that is read out by the laser beam's
phase shift. Although this fact is a direct consequence of the formalism
developed in this paper, we would like to give an intuitive example 
in order to emphasize this point. 

Consider,  for the sake of simplicity,
 a one dimensional elastic test mass
with
two identical defects A and B, 
as shown on Fig.\ 1; A is 
closer to the beam spot than B. Each of these defects creates a random
 stress which pushes apart or pulls together the left and right (relative to the defect)
parts of the test mass. By conservation of momentum, the part of the test mass which is lighter 
 will respond more to the random stress than the other part; therefore defect A
will have a larger effect on the optical readout than the  B. 

Note that if the defects A and B
are positioned symmetrically with respect to the centre of the test mass, they will
have the same effect on the $Q$'s of all elastic modes (we assume for simplicity
that only one-dimensional longitudinal modes are present---and all of them are either
symmetric or antisymmetric with respect to the centre). Therefore, the normal-mode decomposition
applied to the test-mass with just one defect---A or B---would give the same result for the thermal 
noise as read by the laser. Clearly, we have found yet another illustration of the breakdown of the normal-mode
decomposition .

The considerations presented above lead to the following advice
 for real experiments:
 keep the neighbourhood  of the laser beam spot as clean of defects as possible.

Not only does our direct applicatin of the Fluctuation-Dissipation Theorem
have broader validity than the normal-mode decomposition; 
it is also be computationally simpler. In the case of homogeneous structural
damping it yields a simple analytical expression for the
internal thermal noise spectrum [cf. Eq. (\ref{eq:tn3})]:
\begin{equation}
S_x(f)={4k_{B} T\over f}{1-\sigma^2\over \pi^3 E_0 r_0} I \phi\left[1-O\left({r_0\over R}\right)\right].
\label{eq:tn4}
\end{equation}
This result is consistent with the numerical sum-over-modes done in Ref. \cite{raab}
and is accurate when the radius of the laser beam is small relative
to the size of the test mass, i.e. in the regime when the sum over modes converges
especially slowly. When $r_0/R$ is not small, a numerical solution of the elasticity
equations to deduce the dissipation power $W_{\rm diss}$, and thence the thermal noise
(10), is straightforward and is probably also much simpler than performing a sum
over modes.  

\section{acknowledgement}
This work would not have been possible without
discussions  and help from Vladimir Braginsky, Ron Drever,
Darrell Harrington, Nergis Mavalvala, Fred Raab, Glenn Soberman and Kip Thorne.
In particular, Glenn Soberman suggested the method of integration
in Eq. (\ref{eq:integral}), and Kip Thorne carefully reviewed the manuscript
and made a few significant corrections and suggestions.
This work was supported in part by  NSF grant PHY-9424337.

\begin{appendix}
\section{The strain energy in a test mass subjected to a gaussianly distributed 
surface pressure}
The objective of this Appendix is to derive Eq. (\ref{eq:umax1})
of Section III for the energy of  elastic strain
in a cylindrical test mass when the pressure $P(\vec{r})=F_0 f(\vec{r})$
is applied to one of it's circular faces. (As was discussed in Section III, we can assume
that the pressure is constant in time since  LIGO's detection frequencies are much lower
than the lowest normal-mode frequency). For a circular laser beam
with a gaussian intensity profile $f(\vec{r})$ is given by [cf. Eq. (\ref{eq:formfactor})] 
\begin{equation}
f(\vec{r})={1\over \pi r_0^2}e^{-r^2/r_0^2},
\label{eq:formfactor1}
\end{equation} 
where we assume that the centre of the light spot coincides with the centre
of the test mass's circular face.

If the radius of the laser beam $r_0$ is small compared to the size of 
the test mass, we can approximate the test mass by an infinite elastic 
half-space. Then our calculation of the elastic energy is correct up to a
fractional accuracy  of $O(r_0/R)$, where $R$ is the characteristic size of the test mass.

Let $y(\vec{r})$ be the normal displacement of the surface at location 
$\vec{r}$ under the action  of the pressure $P(\vec{r})$. In the linear approximation
of small strains
\begin{equation}
y(\vec{r})=\int G(\vec{r},\vec{r^{\prime}}) P(\vec{r^{\prime}}) d^2 r^{\prime},
\label{eq:dy}
\end{equation}
where $G(\vec{r}, \vec{r^{\prime}})$ is a Green's function. The calculation
of $G$ is a non-trivial albeit standard 
  exercise in elasticity theory \cite{landau},
which gives
\begin{equation}
G(\vec{r}, \vec{r^{\prime}})={1-\sigma^2\over \pi E_0}{1\over |\vec{r}-\vec{r^{\prime}}|},
\label{eq:green}
 \end{equation}
where $\sigma$ is the Poisson ratio and $E_0$  the Young's modulus of the material.
The elastic energy stored in the material is 
\begin{eqnarray}
U_{\rm max}&=&\int P(\vec{r})y(\vec{r}) d^2 r\nonumber\\
           &=&{1-\sigma^2\over \pi E_0}\int {P(\vec{r})P(\vec{r^{\prime}})\over
              |\vec{r}-\vec{r^{\prime}}|} d^2 r d^2 r^{\prime}\label{eq:integral}\\
           &=&{1-\sigma^2\over \pi^3 E_0 r_0^4} F_0^2 \int
              {e^{-({r^2+{r^{\prime}}^2)/ r_0^2}}\over\sqrt{r^2+{r^{\prime}}^2-2rr^{\prime}\cos \theta}}
              d^2 r d^2 r^{\prime} ,\nonumber
\end{eqnarray}
where $\theta$ is the angle between $\vec{r}$ and $\vec{r^{\prime}}$.  
The integral in the last term of Eq. (\ref{eq:integral}) (as was pointed
out by Glenn Sobermann)  can be taken by 
introducing ``polar'' coordinates $R$ and $\phi$: $r=R\cos\phi$, $r^{\prime}=R\sin\phi$.
One then integrates out the radial part of the integrand and expands the remaining angular
part in a power series with respect to $\cos\theta$; termwise integration of this power
series finally yields Eq. (\ref{eq:umax1}) [up to a fractional error of $O(r_o/R)$]
\begin{equation}
U_{\rm max}\simeq {F_0^2\over\pi^2 E_0 r_0}(1-\sigma^2)I,
\label{eq:umax2}
\end{equation}
where
\begin{equation}
I={\pi^{3/2}\over 4}\left[1+\sum_{n=1}^{\infty}{(4n-1)!!\over (2n)! 4^n (2n+1)}\right]\simeq 1.87322.
\label{eq:I}
\end{equation}
It can be shown that if, instead of an infinite half-space, we 
consider a finite cylindrical test mass, the leading
fractional correction to the elastic energy is of the order $O(r_0/R)$.

\end{appendix}

\newpage
\begin{figure}
\caption[]{
Identical defects A and B create fluctuating strees in different parts
of the test mass. The stress created by defect A will influence 
the phase shift of the laser beam readout more than the stress created by 
defect B, although both A and B make identical contributions to 
$Q$'s  of the test mass's elastic modes.
}
\label{fig:readout}
\end{figure}
\end{document}